 \documentclass[aps,pra,showpacs]{revtex4}
 \usepackage{graphicx}

\begin{document}

 \def\BE{\begin{equation}}
 \def\EE{\end{equation}}
 \def\BA{\begin{array}}
 \def\EA{\end{array}}
 \def\DS{\displaystyle}
 \def\BEA{\begin{eqnarray}}
 \def\EEA{\end{eqnarray}}
 \def\vp{\vec\rho}
 \def\a{\alpha}
 \def\b{\beta}
 \def\l{\lambda}
 \def\vq{\vec q}
 \def\Om{\Omega}
 \def\O{\Omega}
 \def\om{\omega}
 \def\sinc{{\,\rm sinc\,}}
 \def\bP{\textit{\textbf{P}}}
 \def\bX{\textit{\textbf{X}}}
 \def\k{\kappa}
 \def\r{\vec\rho}
 \def\q{\vec q}

\title{A quantum volume hologram}

 \author{Denis~V.~Vasilyev$^{1}$, Ivan~V.~Sokolov$^{1}$,
and Eugene~S.~Polzik$^{2}$}
 \affiliation{1 V.~A.~Fock Physics Institute, St.~Petersburg State University,
 198504 Petrodvorets, St.~Petersburg, Russia}
 \affiliation{2 QUANTOP, Danish Research Foundation Center for Quantum Optics,
 Niels Bohr Institute, DK 2100 Copenhagen, Denmark}

 \begin{abstract}
We propose a new scheme for parallel spatially multimode quantum memory for
light. The scheme is based on counter-propagating quantum signal wave and
strong classical reference wave, like in a classical volume hologram, and
therefore can be called a {\sl quantum volume hologram}. The medium for the
hologram consists of a spatially extended ensemble of atoms placed in a
magnetic field. The write-in and read-out of this quantum hologram is as simple
as that of its classical counterpart and consists of a single pass
illumination. In addition we show that the present scheme for a quantum
hologram is  less sensitive to diffraction and therefore is capable of
achieving higher density of storage of spatial modes as compared to previous
proposals. A quantum hologram capable of storing entangled images can become an
important ingredient in quantum information processing and quantum imaging.
 \end{abstract}


 \maketitle

 \section{introduction}

Quantum memory for light is an essential part of quantum information protocols,
such as quantum repeaters, distributed quantum computation, and quantum
networks. A number of approaches based on storage in atomic  ensembles were
developed recently, including quantum non-demolition (QND) interaction,
electromagnetically induced transparency (EIT), Raman interaction and photon
echo. For a comprehensive recent review on quantum interfaces between  light
and matter see \cite{Hammerer08}. Multimode quantum memories are in the center
of current research due to their potential for enhanced storage capacity and
state purification and error correction,e.g., in quantum repeaters
\cite{Simon07}.

Spatially--multimode parallel quantum protocols for light without memory have
been central for the field of quantum imaging. Examples are quantum holographic
teleportation \cite{Sokolov01,Gatti04} and telecloning \cite{Magdenko07}, and
quantum dense coding of optical images \cite{Golubev06}. The spatially
multimode light in an entangled Einstein - Podolsky - Rosen (EPR)  quantum
state \cite{Kolobov99} has been recently experimentally demonstrated via
four-wave mixing in \cite{Lett08}.

A quantum memory protocol based on a QND interaction, for the single--mode case
experimentally demonstrated in \cite{Julsgaard04}, achieves storage in the
ground state of an ensemble of spin--polarized  atoms. An extension of the
QND--based scheme of quantum memory to spatially multimode configuration, able
to store quantum images, has been proposed recently in \cite{Vasilyev08}. In
this proposal two passes of light are required to store, and another two -- to
retrieve quantum state of an image from the hologram. In addition, perfect
squeezing of the initial state of light and/or atoms is required for an ideal
performance of the holographic memory.

Another approach to the multimode quantum memory based on
phase-matched backward propagation retrieval out of the EIT and the
Raman type memories has been recently proposed for the multiple
frequency-encoded qubits storage in a single ensemble
\cite{Shurmacz08}. A gradient echo memory has been proposed for
storage of several frequency encoded modes \cite{Sellars08}.

The optical image storage has been demonstrated in recent experiments
\cite{Camacho07,Shuker08,Vudyasetu08}. There was observed \cite{Camacho07} a
several nanoseconds delay of optical pulses which contain on average less than
one photon and carry two-dimensional images. The delay has been achieved by
using slow light in an atomic ensemble. However, preservation of quantum field
properties was not demonstrated in the paper. The storage of classical images
in a warm atomic vapor by EIT interaction has been investigated in
\cite{Shuker08,Vudyasetu08}.

In this paper we present a new scheme of the multimode quantum memory which we
call a quantum volume hologram. Our proposal makes use of two concepts. The
first one stems from a volume hologram, proposed by Denisyuk \cite{Denisyuk62}
for classical storage of optical images. The volume hologram is written by the
counter-propagating signal and reference waves. There are two sublattices,
produced by the waves interfering in the medium, each of them stores one
quadrature of the signal field. Since both quadratures are stored, there  are
no virtual and real images during the readout, by contrast to a thin hologram.

The second concept comes from \cite{Muschik06} where it was shown that a
combination of a constant magnetic field with the QND interaction allows to
couple two components of the collective atomic spin used for the quantum memory
to two quadratures of light.

In the present paper we  propose to write volume hologram (which implies the
counter-propagating geometry) onto an atomic ensemble with spins rotating in a
constant magnetic field. Hence, in our model all degrees of freedom of atoms and light
are symmetrically involved in the interaction. We show that an input state of
spatially--multimode light can be written onto the quantum volume hologram in a single
pass. A sideband (with respect to the carrier frequency of the strong field $\omega$)
$+\O$ spectral component of light is written onto the collective spin coherence wave,
which propagates in the medium with a certain phase velocity.

Our analysis shows, that  the volume hologram for the ground state spin
$J=1/2$, when considered in the limit of a single spatial mode, has features
similar to those described in \cite{Nunn07} for $\Lambda$--scheme of atomic
levels (possible only for $J>1/2$). It appears that an effective increase in
the number of the system degrees of freedom due to the counter-propagating
geometry in combination with the spin rotation has an effect similar to the
increased complexity of the atomic level structure in the $\Lambda$--scheme
based memories.

The paper is organized as follows. First, we discuss the basics of the model
and derive the light and matter equations of motion in the paraxial
approximation using averaging over fast oscillations in space and time. Next,
we consider transfer of an input multimode quantum state encoded in the signal
wavefront at the write and readout stages. Finally, the stored transverse modes
number is estimated for the volume and the thin \cite{Vasilyev08} quantum
hologram and we come to conclusion.

 \section{Single pass volume hologram with spatial resolution}

\begin{figure}
 \begin{center}
 \includegraphics[width=80mm]{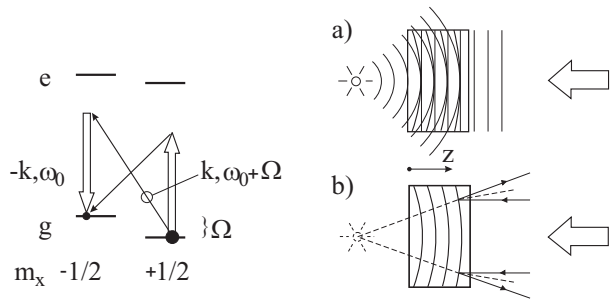}
 \caption{Schematic of the level structure and of the write (a) and
 readout (b) stages of quantum volume hologram}
 \label{fig1_Wright_readout}
 \end{center}
 \end{figure}

The scheme, illustrating our quantum memory protocol is shown in
Fig.~\ref{fig1_Wright_readout}.
We consider an ensemble of motionless atoms which we for simplicity assume to
have an angular momentum of $J=1/2$ both in the ground and in the excited
states, located at random positions. The long-lived ground state spin of an
atom ${\vec J}^a$ is initially oriented along the constant magnetic field in
the vertical direction $x$. The atomic spins rotate around the vertical axis
with a circular frequency $\Om$. A classical off-resonant $x$-polarized plane
wave at frequency $\om_0$ with a slowly varying amplitude $A_x$ (assumed to be
real) propagates in $-z$ direction. The input signal wave is a weak quantized
$y$-polarized field at the same frequency $\omega_0$, propagating in $+z$
direction. In what follows we consider this multimode input field with a slowly
varying amplitude $A_y(\vec r,t)\ll A_x$ in the paraxial approximation.

In order to construct the Hamiltonian of the system we first consider an atomic
layer with the thickness (in $z$ direction) smaller than $\lambda$. Within this
layer the phase difference between the signal field and the driving wave is
constant. Hence, the interaction within the slice is described by the well
known quantum non-demolition (QND) Hamiltonian \cite{Hammerer08}. The QND
light-matter interaction in the each layer leads to two basic effects: (i) the
Faraday rotation of light polarization due to longitudinal quantum
$z$-component of collective atomic spin of the slice; and (ii) the atomic spin
rotation, caused by the unequal light shifts of the ground state sub-levels
with $m_z=\pm 1/2$ in the presence of quantum fluctuations of circular  light
polarizations. The relevant part of the Hamiltonian is \cite{Hammerer08}:
 \BE
H = \frac{2\pi k_0 |d|^2}{\om_{eg}-\om_0}\int_V d\vec r \sum_a
J_z^a(t) S_z(\vec r,t)\delta(\vec r -\vec r_a).
 \label{interaction}
 \EE
Here $\om_{eg}$ is the frequency of the atomic transition, $d$ is
the dipole matrix element, and $k_0 =\om_0 c$. For the
counterpropagating  signal and driving waves, the $z$-component of
the Stokes vector  $S_z(\vec r,t) = 2A_x\,Im\left[A_y(\vec
r,t)e^{2k_0z}\right]$ in (\ref{interaction}) rapidly (i.e. on the
scale $\lambda$) oscillates along $z$. The
amplitude $A_y$ is defined via
 $$
 A_y(z,\r,t)=\int\frac{dk_z}{2\pi}\int\frac{d\vec
q}{(2\pi)^2}\sqrt{\omega(k)/k_0}\,a_y(\vec k)
 \exp[i(\q\cdot\r+(k_z-k_0)z-(\omega(k)-\omega_0)t)],
 $$
here $a_y(\vec k)$ and $a_y^\dagger(\vec k)$ are the annihilation
and creation operators for the wave $\vec k$, which obey standard
commutation relations $ [a_y(\vec k),a_y^\dagger(\vec
k\,')]=(2\pi)^3\delta(\vec k-\vec k\,')$, $[a_y(\vec  k),a_y(\vec
k\,')]=0$. By using these commutation relations in the paraxial
approximation, one finds \cite{Kolobov99} the commutation relation
for the slowly varying amplitude of the quantized signal field
$A_y$,
 \BE
 [A_y(z,\r,t),A_y^\dag(z',\r\,',t)] \equiv c\tilde{\delta}(\vec r -
\vec r\,')\approx
 \label{Ay_commutator}
 \EE
 $$
 c\left(1-\frac{i}{k_0}\frac{\partial}{\partial
 z}-\frac{1}{2k^2_0}\nabla^{\,2}_{\bot}\right)\delta(\vec r - \vec r\,'),
 $$
where $\vec r=(\r,z)$, $\vec k=(\q,k_z)$. We do not consider here
the $y$-polarized quantized field co-propagating with the driving
wave because its evolution is independent of the signal wave under
consideration.

We introduce the density of the collective spin as $\vec J(\vec r)
= \sum_a \vec J^a \delta(\vec r-\vec r_a)$. The averaged over
random positions of the atoms commutation relation for the $y$,
$z$ components of the collective spin is
 $$
\overline{[J_y(\vec r),J_z(\vec r\,')]} = i\sum_a \langle J_x^a \rangle
\overline{\delta(\vec r-\vec r_a)\delta(\vec r\,'-\vec r_a)}^a =
i n_a \langle J_x^a \rangle \delta(\vec r-\vec r\,').
 $$
Here $n_a$ is the average density of atoms. The field-like
variable for the spin subsystem,
 $$
B(\vec r,t)= \{J_y(\vec r,t)+iJ_z(\vec r,t)\}/\sqrt{2n_a \langle
J_x^a \rangle},
 $$
obeys the standard boson commutation relation:
 \BE
 \overline{[B(\vec r,t),B^{\dag}(\vec r\,',t)]} =  \delta(\vec r-\vec r\,').
 \label{B_commutator}
 \EE

In view of the fact that the atoms are prepared in the spin up
state, the operator $B^{\dag}$ can be considered as the creation
operator for atoms in the spin down state, or the collective
projection operator $|+1/2\rangle \to |-1/2\rangle$.

The full Hamiltonian of our model includes the energy of a free
electromagnetic field, the interaction of the atomic spin with
constant magnetic field and the effective Hamiltonian of the QND
interaction. The Hamiltonian reads,
 \BE
H = \int_V d\vec r\left\{ \frac{\hbar
\omega_0}{c}A_y^\dag(z,\vp,t)\,A_y(z,\vp,t) + \hbar\Om
\,B^\dag(z,\vp,t)\,B(z,\vp,t) - \right.
 \label{full_hamiltonian}
 \EE
 $$
\left. \frac{\sqrt{2}\pi k_0 |d|^2}{\om_{eg}-\om_0} \sqrt{n\langle
J_x^a\rangle}A_x(z,t) \big[B(z,\vp,t)-h.c.\big]
\big[A_y(z,\vp,t)e^{i2k_0z}-h.c.\big]\right\}.
 $$
We describe the evolution of our system in the Heisenberg picture.
Using the commutation relations (\ref{Ay_commutator}) and
(\ref{B_commutator}) for the field and atomic variables, after
simple transformations we obtain:
 \BE
 \label{A_evoution}
\left(\frac{\partial}{\partial
z}-\frac{i}{2k_0}\nabla^{\,2}_{\bot}
+\frac{1}{c}\frac{\partial}{\partial t}\right) A_y(z,\vp,t) =
\frac{\k}{\sqrt{LT}}\,{\rm Im}B(z,\vp,t)e^{-i2k_0z},
 \EE
 \BE
 \label{B_evolution}
\frac{\partial}{\partial t} B(z,\vp,t) = -i\Om B(z,\vp,t) +
\frac{\k}{\sqrt{LT}}{\rm Im}\left[A_y(z,\vp,t)e^{i2k_0z}\right].
 \EE
Here $T$ is the duration of the flat-top pulse of the driving field
and $L$ is the atomic cell length. The dimensionless coupling
constant
 \BE
\k=\frac{2\pi k_0 |d|^2}{\hbar (\om_{eg}-\om_0)}\sqrt{2n_a\langle
J_x^a\rangle A_x^2 LT},
 \EE
should be of the order unity for the memory to work.  The coupling constant can
be written as $\kappa^2=\alpha_0\eta$, where $\alpha_0$ is the resonant optical
depth and $\eta$ is the probability of spontaneous emission \cite{Hammerer08}.
Since $\eta\ll 1$ is required in order to neglect the effect of spontaneous
emission from the excited atomic level, the usual condition for an efficient
quantum interface $\alpha_0 = \lambda^2n_aL/2\pi\gg1$ should be fulfilled. We
introduce the Fourier transform via
 \BE
a(z,\vq,t) = \int d\vp\,A_y(z,\vp,t) e^{-i\vq \, \vp},
 \EE
and similar for atomic variables, and arrive at the set of basic equations in
the Fourier domain,
 \BE
 \label{a_evolution}
\left(\frac{\partial}{\partial z}+
i\frac{\vq\,^2}{2k_0}+\frac{1}{c}\frac{\partial}{\partial t}\right)
a(z,\vq,t) =
\frac{\k}{\sqrt{LT}}\frac{1}{2i}\left[b(z,\vq,t)-b^\dag(z,-\q,t)\right]
e^{-i2k_0z},
 \EE
 \BE
\frac{\partial}{\partial t} b(z,\vq,t) = -i\Om b(z,\vq,t) +
\frac{\k}{\sqrt{LT}}\frac{1}{2i}\left[a(z,\vq,t)e^{i2k_0z} -
a^\dag(z,-\vq,t)e^{-i2k_0z}\right].
 \label{b_evolution}
 \EE
The considered above field and atomic amplitudes are rapidly oscillating at
frequency of the order $\O$. In what follows we introduce the slowly varying
amplitudes of collective spin $b_{\pm k}(z,\q,t)$ in the rotating at $\O$
frame, and slowly varying amplitudes $a_{\pm\O}(z,\q,t)$ for the signal field
sidebands at frequencies $\omega_0\pm\Omega$.

Next, we should take into account that the field amplitudes are defined as
slowly varying along the $z$ axis, but the spin amplitudes are not, as seen
from (\ref{b_evolution}). The fast modulation of the collective spin at the
longitudinal spatial frequency $2k_0$ is just a consequence of the counter
propagating geometry of the volume hologram. The thin atomic slices discussed
above have a length of the order of a fraction of $\lambda$. This imposes
limitations on the atomic motion during the storage time: the atoms should not
transport coherence to another slice. A solid state, an optical lattice, or an
ultra cold atomic ensemble should be used  to fulfil this condition. The slow
in space and time variables look like
 \BE
a_{\pm\Omega}(z,\vq,t) = a(z,\vq,t)e^{\pm i\Om t},
 \label{slow_ab}
 \EE
 $$
b_{\pm k}(z,\vq,t) = b(z,\vq,t)e^{i(\Om t \mp 2k_0z)}.
 $$
Quantum memory is realized via the dynamical coupling between
$a_{\Omega}(z,\vq,t)$ and $b_{k}(z,\vq,t)$. We insert these
variables into (\ref{a_evolution}), (\ref{b_evolution}) and
perform  averaging on a time scale $1/\Omega \ll t_{av} \ll T$ and
on a spatial scale in $z$ direction $\lambda \ll z_{av} \ll L$.
This implies that the pulse duration and the cell length are taken
large enough, $\Omega T \gg 1$ and $k_0 L \gg 1$, where the last
condition is typical also for classical volume holograms.

The equations of motion read,
 \BE
\frac{\partial}{\partial z} a_{\Om}(z,\vq,t) =
\left[-\frac{1}{c}\frac{\partial}{\partial t} + i\left(
\frac{\Om}{c} -\frac{\vq\,^2}{2k_0}\right)\right]a_{\Om}(z,\vq,t)
-i\frac{\k}{2\sqrt{LT}}b_k(z,\vq,t),
 \label{a_slow_evolution}
 \EE
 \BE
 \frac{\partial}{\partial
 t} b_k(z,\vq,t) = -i\frac{\k}{2\sqrt{LT}}a_{\Om}(z,\vq,t).
 \label{b_slow_evolution}
 \EE
In what follows we neglect the retardation effects. Both the pulse
length in space, estimated as $c/\delta\Omega$, where
$\delta\Omega$ is the signal spectral width, and the spatial
length $c/\Omega$ of modulation at frequency $\Omega$ are assumed
to be much larger than the cell length, hence $\delta\Omega/c \ll
\Omega/c \ll 1/L$. This allows to omit the terms $\sim 1/c$ on the
right side of (\ref{a_slow_evolution}) as compared to the term
$\sim  b /\sqrt{LT}$, which is estimated  as $\sim a/L$. Here we
make use of (\ref{b_slow_evolution}) and find for typical values
$\k\sim 1$ and $t\sim T$ an estimate $b\sim \sqrt{T/L}\,a$.

Next, we go over to the amplitudes
 \BE
 \label{tilde_amplitudes}
 \tilde a_\O(z,\q,t)= a_\O(z,\q,t) \exp\left(i\frac{q^2}{2k_0} z\right), \qquad
 \tilde b_k(z,\q,t)= b_k(z,\q,t) \exp\left(i\frac{q^2}{2k_0} z\right),
 \EE
and arrive at the equations
 \BE
 \label{Raman_equations}
\frac{\partial}{\partial z} \tilde a_{\Om}(z,\q,t) =
-i\frac{\k}{2\sqrt{LT}}\,\tilde b_k(z,\q,t), \qquad
\frac{\partial}{\partial
 t}\tilde b_k(z,\q,t) = -i\frac{\k}{2\sqrt{LT}}\,\tilde a_{\Om}(z,\q,t).
 \EE
As compared to the equations, previously discussed (e.g.,
\cite{Mishina06,Nunn07} and references therein) for the $\Lambda$--scheme
Raman--type models of spatially single mode quantum memory, we do not encounter
in (\ref{Raman_equations}) the deleterious effect of the Stark shift of atomic
levels by a classical control pulse. The  level shift makes the phase matching
and evolution more complicated, especially for a time--dependent profile of the
control field. This simplification is due to the more symmetrical nature of the
QND interaction.

It follows from (\ref{Raman_equations}) that in our parallel
quantum memory, based on the volume hologram, the signal and the spin
coherence waves with an arbitrary $\q$ (in paraxial approximation)
are interacting like the waves propagating along $z$ direction in a
spatially single mode Raman--type memory. The diffraction does not
modify the state exchange between light and matter, and our memory
is able to store as many orthogonal spatial mode with different
$\q$ as one can fit into a hologram of transverse area $S$. We take
$\Delta q_{x,y} \sim 2\pi/\sqrt{S}$, $|\q| \ll 2\pi/\lambda$, and
assume that the number of transverse modes is much less than $4S/\lambda^2$.

The solution in terms of the amplitudes (\ref{tilde_amplitudes})
is found as an extension to the spatially multimode case of the
results previously obtained for the Raman--type model,
 \BE
\tilde a_{\Om}(z,\vq,t) = \tilde a_{\Om}(0,\vq,t)-
\frac{\k}{2\sqrt{LT}}\int_0^t dt'\sqrt{\frac{z}{t-t'}}\,
J_1\left(\k\sqrt{\frac{z(t-t')}{LT}}\right)\tilde
a_{\Om}(0,\vq,t')
 \EE
 $$
- i\frac{\k}{2\sqrt{LT}}\int_0^z
dz'\,J_0\left(\k\sqrt{\frac{(z-z')t}{LT}}\right)\tilde
b_k(z',\vq,0),
 $$
 \BE
\tilde b_k(z,\vq,t) = \tilde b_k(z,\vq,0) -
\frac{\k}{2\sqrt{LT}}\int_0^z  dz'\sqrt{\frac{t}{z-z'}}\,
J_1\left(\k\sqrt{\frac{(z-z')t}{LT}}\right)\tilde b_k(z',\vq,0)
 \EE
 $$
- i\frac{\k}{2\sqrt{LT}}\int_0^t
dt'\,J_0\left(\k\sqrt{\frac{z(t-t')}{LT}}\right) \tilde
a_{\Om}(0,\vq,t').
 $$

Consider the dimensionless coordinates and amplitudes,
 \BE
 \label{dimless}
\frac{z}{L} = \xi, \quad \frac{t}{T} = \tau, \quad a_\O(z,\q,t)\sqrt{T} =
\alpha_\O(\xi,\q,\tau), \quad b_k(z,\q,t)\sqrt{L} = \beta_k(\xi,\q,\tau).
 \EE
Here $\a(z,\r,t)^\dag \a(z,\r,t)$ corresponds to the number of signal photons
per $cm^2$ of the beam cross section during the interaction time $T$, and
$\b(z,\r,t)^\dag \b(z,\r,t)$ gives the number of flipped spins  per $cm^2$ of
the hologram cross section at the length $L$. The equations above become
 \BE
 \label{out_light}
\tilde\a_{\Om}^{(out)}(\vq,\tau) = \int_0^1 d\tau'
G_1(\tau-\tau',1)\tilde\a_{\Om}^{(in)}(\vq,\tau') -i \int_0^1 d\xi'
G_0(1-\xi',\tau)\tilde
\b_k^{(in)}(\xi',\vq),
 \EE
 \BE
\tilde{\b}_k^{(out)}(\xi,\vq) = \int_0^1 d\xi' G_1(\xi-\xi',1)\tilde
\b_k^{(in)}(\xi',\vq) -i \int_0^1 d\tau'
G_0(1-\tau',\xi)\tilde \a_{\Om}^{(in)}(\vq,\tau'),
 \label{out_spins}
 \EE
where $\tilde\a_{\Om}^{(in)}(\vq,\tau)=\tilde\a_{\Om}(0,\vq,\tau)$,
$\tilde\a_{\Om}^{(out)}(\vq,\tau)=\tilde\a_{\Om}(1,\vq,\tau)$, and
$\tilde{\b}^{(in)}(\xi,\vq)=\tilde{\b}_k(\xi,\vq,0)$,
$\tilde{\b}^{(out)}(\xi,\vq)=\tilde{\b}_k(\xi,\vq,1)$.  The integral
kernels are given by
 \BE
G_0(p,q)= \frac{\kappa}{2}J_0(\kappa \sqrt{pq}), \quad
G_1(p,q)=\delta(p)-\frac{\kappa}{2}J_1(\kappa\sqrt{qp})\sqrt{q/p}\, \theta(p),
 \EE
with the $n$th Bessel function of the first kind denoted by $J_n$.

Similar to the analysis of
\cite{Nunn07}, we make use of the fact that the kernels $G_{0,1}$
share eigen functions,
 \BEA
 G_0(1-x,y) &=& \sum_i\phi_i(y)\lambda_i\phi_i(1-x),\\
 G_1(y-x,1) &=& \sum_i\phi_i(y)\mu_i\phi_i(1-x),\nonumber
 \label{eugen_modes}
 \EEA
and their eigen values satisfy the constraint
$\lambda_i^2+\mu_i^2=1$ for all $i$. By applying the variables
decomposition of the form
 \BEA
 \tilde\a^{(in)}_\O(\q,\tau) &=& \sum_j \tilde\a^{(in)}_{\O,j}(\q)\phi_j(1-\tau),\\
 \tilde\b^{(in)}_k(\xi,\q)  &=& \sum_j \tilde\b^{(in)}_{k,j}(\q)\phi_j(1-\xi),\nonumber
 \label{signal_decomposition}
 \EEA
one arrives at the input--output transformation for one cycle of the
light--matter interaction:
 \BEA
 \label{in_out_relation}
\tilde\a^{(out)}_\O(\q,\tau) &=& \sum_i \left[\mu_i\tilde\a^{(in)}_{\O,i}(\q) -i
\lambda_i\tilde\b^{(in)}_{k,i}(q)\right]\phi_i(\tau), \\
 \tilde\b^{(out)}_k(\xi,\q) &=& \sum_i
\left[-i\lambda_i\tilde\a^{(in)}_{\O,i}(\q) +
\mu_i\tilde\b^{(in)}_{k,i}(q)\right]\phi_i(\xi).\nonumber
 \EEA
The transformation is of a beamsplitter type, it is unitary and hence
respects the correct bosonic commutation relations. The Eq.
(\ref{in_out_relation}) shows, that if the light signal stored in an
atomic hologram has the temporal profile of an eigen mode of $G_0$ with
the eigen value $\lambda_i$ close to unity, the initial state of the atomic spins
in the mode $i$ in erased since the
corresponding eigen value $\mu_i$ is close to zero.

\begin{figure}
 \begin{center}
 \includegraphics[width=70mm]{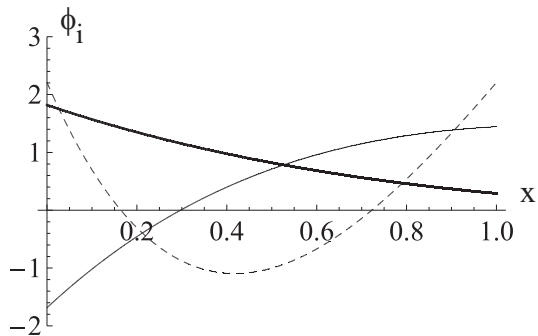}
 \caption{The first three eigen functions: $\phi_1$, $\phi_2$, and $\phi_3$
 (the bold, thin and dashed line respectively) for $\kappa=4$. The eigen values
 are $\lambda_1\approx0.988$,  $\lambda_2\approx -0.518$, and $\lambda _3\approx0.043$.}
 \label{fig2_Eigen_functions}
 \end{center}
 \end{figure}

In order to read out the image stored in the hologram, one has to repeat the
procedure and pass another classical light pulse through
the atoms. To describe the whole write--readout cycle
of parallel quantum memory, we combine the transformations
 $
\left\{{\tilde\a_\O}^{W(in)}(\q,\tau),\,
{\tilde\b_k}^{W(in)}(\xi,\q)\right\} \to {\tilde\b_k}^{W(out)}(\xi,\q)
 $
for the write stage, and
 $
\left\{{\tilde\a_\O}^{R(in)}(\q,\tau),\, {\tilde\b_k}^{R(in)}(\xi,\q) =
{\tilde\b_k}^{W(out)}(\xi,\q)\right\} \to {\tilde\a_\O}^{R(out)}(\q,\tau)
 $
for the readout stage, and restore the diffraction factor (see
(\ref{tilde_amplitudes})). This yields,
 \BE
\a_\O^{R(out)}(\q,\tau) = \exp\left(-i\frac{q^2}{2k_0}L\right)
\sum_i\left\{\mu_i \a_{\O,i}^{R(in)}(\q) -i\lambda_i\sum_j f_{ij}
\left[\mu_j\tilde\b_{k,j}^{W(in)}(\q) -i\lambda_j\a_{\O,j}^{W(in)}(\q)
\right]\right\}\phi_i(\tau),
 \label{whole_cycle}
 \EE
where
 \BE
 f_{ij} = \int_0^1 d\xi \phi_i(1-\xi)\phi_j(\xi),
 \EE
is the eigen functions overlap factor. In Fig.\ref{fig2_Eigen_functions}
we plot the first eigen functions evaluated numerically.
The Eq.(\ref{whole_cycle}) shows that the input  (at the write stage) quantized amplitude
of the signal field in the $i$th eigen mode gets restored at $z=L$ with the factor $\sim
\exp\left(-iq^2 L/2k_0\right)\lambda_i^2 f_{ii}$. Like in the Raman--type schemes, our
model demonstrates the time reversal symmetry between the input and output modes, as
follows from the input (\ref{signal_decomposition}) and output (\ref{whole_cycle}) signal
decomposition. The read-in coupling parameter $\kappa\approx 4$ and a read-out
coupling parameter in excess of 20 is required to achieve memory efficiency
$\geq0.95$ \cite{Nunn07}.

The storage capacity, that is the number of transverse modes that the  quantum
hologram can store is one of the most important parameters for the memory
applications. The memory capacity of the previously proposed \cite{Vasilyev08}
thin quantum hologram is strongly limited by diffraction. Namely, the
diffraction spread of an input image element of the linear size $d$ should be
small, $\lambda/d\cdot L\leq d$. Hence, the number $N$ of the resolved by the
thin hologram image elements (modes) is determined by the sample Fresnel number
$F_{N}$, so that $N \sim S/d^2 \sim S/\lambda L=F_{N}$, where $S$ is the sample
cross section. The authors of \cite{Shurmacz08} who have considered the
multiple optical modes storage in an atomic ensemble with appropriately phase-matched backward propagating retrieval also estimate the number of stored
transverse modes to be equal to $F_{N}$.

In comparison, the effect of diffraction on the signal wavefront stored in the
volume hologram is the same as in free space and can be compensated simply by
using a lens system with the input focal plane at $z=0$. The limitation on the
capacity for the volume hologram comes first from the paraxial approximation
$\lambda/d\leq\varepsilon$, where $\varepsilon \ll 1$ is the small parameter of
the paraxial approximation. The second limitation is due to the geometry of the
sample, $\lambda/d\cdot L\approx \sqrt{S}$, so that the quantized waves
propagate inside the atomic cell. Hence the number of stored modes in the
volume hologram is estimated to be  $S/d^{2} \sim {\rm min}\{\varepsilon^2
S/\lambda^2,\,F_{N}^{2}\}$. For not too elongated samples with $\sqrt{S}/L\geq
\varepsilon$ the number of modes is equal to $\varepsilon^2 S/\lambda^2$ which
is larger than for memories proposed so far. There are indications that for the
co-propagating write-in and read-out $\Lambda$-- scheme based multimode
memories the capacity is also of the order of $\varepsilon^2 S/\lambda^2$
\cite{Sorensen09}.

 \section{Conclusions}

We have presented an extension of the classical volume hologram into the
quantum domain. The volume quantum hologram can store entangled and other quantum images and offers a new approach to a
parallel spatially multimode quantum memory for light. The equations of motion for the
quantized field and spin coherence waves are derived and examined in paraxial
approximation. Our analysis could be of use for the spatially--multimode
generalization of other quantum memory models.

The counter--propagating geometry of the classical and quantized fields,
combined with the spins rotation in a constant magnetic field, allows for the
efficient quantum image transfer between light and matter in a single pass of
light. The volume hologram proposed here does not require any extra operations,
such as squeezing, in order to achieve, in principle, perfect performance.

In our scheme, different spatial modes of the incoming field are stored in the
corresponding orthogonal spatial modes of the atomic ensemble. We  show that the
quantum volume hologram is less sensitive to diffraction and is capable of storing
more spatial modes  as compared to the previously proposed \cite{Vasilyev08} thin
hologram in the co--propagating geometry.

Although we considered spin $1/2$ atoms, our analysis can be easily generalized
to an arbitrary angular momentum states in alkali atoms provided that the optical
detuning is greater than the excited state hyperfine structure \cite{Hammerer08}.
Future work on volume quantum holograms will include modified geometries of the
interacting waves which should allow for relaxing of the requirement of motionless
atoms, as well as studies of multimode quantum entanglement between light and matter.\\

We would like to thank Anders S{\o}rensen for fruitful discussions. This research has been funded by the European Commission FP7 under the grant
agreement n° 221906, project HIDEAS. The authors also acknowledge the support
of the Russian Foundation for Basic Research under the projects 05-02-19646,
08-02-00771, and 08-02-92504 (D.V.), and the support by INTAS (Grant 7904).
Part of the research was performed within the framework of GDRE "Lasers et
techniques  optiques de l'information".

\end{document}